# COMPARISON OF MODIFIED DUAL TERNARY INDEXING AND MULTI-KEY HASHING ALGORITHMS FOR MUSIC INFORMATION RETRIEVAL

[1]Rajeswari Sridhar, [2]A. Amudha, [3]S. Karthiga and [4]Geetha T V

Department of Computer Science and Engineering, Anna University, Chennai – 600025
[1] rajisridhar@gmail.com
[2] amudha_be@rediffmail.com
[3] karthiga.s@gmail.com
[4] tvgeedir@cs.annauniv.edu

## ABSTRACT

*In this work we have compared two indexing algorithms that have been used to index and retrieve Carnatic music songs. We have compared a modified algorithm of the Dual ternary indexing algorithm for music indexing and retrieval with the multi-key hashing indexing algorithm proposed by us. The modification in the dual ternary algorithm was essential to handle variable length query phrase and to accommodate features specific to Carnatic music. The dual ternary indexing algorithm is adapted for Carnatic music by segmenting using the segmentation technique for Carnatic music. The dual ternary algorithm is compared with the multi-key hashing algorithm designed by us for indexing and retrieval in which features like MFCC, spectral flux, melody string and spectral centroid are used as features for indexing data into a hash table. The way in which collision resolution was handled by this hash table is different than the normal hash table approaches. It was observed that multi-key hashing based retrieval had a lesser time complexity than dual-ternary based indexing The algorithms were also compared for their precision and recall in which multi-key hashing had a better recall than modified dual ternary indexing for the sample data considered.*

## 1. Introduction.

In this multimedia world, a large collection of digital music has become available [1]. This necessitates efficient management and retrieval of music collection in large music databases. In addition to this, content-based retrieval has gained much attention than mere metadata based retrieval and hence a need for content-based retrieval is the need of the hour [2]. Content-based music retrieval is defined as the process of finding similar occurrences from a music database, when a segment of music piece is given as input. In the meta data based retrieval, similar pieces of music are retrieved when input query is submitted in terms of title, singer or genre. Content based music retrieval requires indexing for all the available music piece in the input and the process of indexing the music piece is performed either manually or automatically. Manual indexing refers to indexing the music piece based on title, singer, genre, query phrase etc and hence it is a cumbersome process. For automatic indexing of songs, in addition to the title, singer, genre, signal level features of the songs are also used for indexing. The features normally chosen can be musical features like rhythm, melody, timbre, intensity, melody string which is represented by their signal features in terms of Mel-Frequency Cepstral Coefficient, spectral flux and centroid or a combination of other features. Hence it is clear that automatic indexing of songs based on features would give a better result in the retrieval since it helps in establishing similarity between songs based on the features. In this paper we have compared two indexing algorithms, the one proposed by us and the modification that has been done by us to an algorithm proposed for western music for indexing and retrieval proposed by Chuan-Wang Chang et al [2]. Hence, here we have compared both the algorithms with reference to automatic indexing and retrieval of Carnatic music songs.

This paper is organized as follows: Section 2 discusses some work in the indexing arena, Section 3 on some basic characteristics of South Indian music which we have considered for indexing, Section 4 discusses about multi-key hashing indexing algorithm, Section 5 discusses on the modification done to the western music algorithm proposed by Chuan-Wang Chang et al to cater to Carnatic music, Section 6 on the Experimental Set up and Results and finally Section 7 concludes the paper

## 2. Existing work

It is evident that automatic content based music information retrieval greatly relies on a good indexing algorithm. The indexing algorithm should be time efficient and in addition to time efficiency should also aid in good precision and recall during the process of retrieval. Several algorithms have been proposed by various researchers for the process of information retrieval. A scalable P2P system for content based music information retrieval has been developed by George Tzanetakis [1]. This algorithm is based on Rendezvous points. The authors have extracted features based on the Short Time Fourier Transform and Mel-Frequency Cepstral Coefficients to represent sound texture, rhythm and pitch content. The means and variances are computed for the Spectral Centroid, Roll-off, Flux and Zero-Crossings and the first 5 Mel Frequency Cepstral Coefficients (MFCC) over a 1 second texture window are calculated for representing Spectral Texture. These features are called as rendezvous points. The authors have designed the algorithm to search based on only a selected quality or qualities of the music piece while ignoring all other parameters. This algorithm had a high time complexity for the process of search and retrieval, since focus was not given on the indexing process to aid in efficient retrieval.

In another algorithm proposed by Andreas Rauber et al, for the automatic indexing of music the genre of the music is used for indexing either directly or through the features conveying the genre of the music piece [3]. This algorithm used time-invariant features which are extracted based on psychoacoustic models for representing the genre of the input music signal. Based on the extracted features a clustering algorithm was used to group similar genre music together which are then used as indices for the given music piece. This algorithm used only the genre parameter for retrieval and hence has to be modified to include other features of the song.

In another work by Hsuan-Huei Shih et al , the authors have assumed that the input file would be in MIDI representation and have modified a table driven algorithm for indexing using the tempo characteristics of the signal [4]. In this work a bar index table is built based on the tempo characteristics and then using the Lempel – Ziv algorithm retrieval is performed based on the bar indexes. This algorithm has given better results when the input is available in MIDI representation.

In another work by Cheng Yang a spectral indexing algorithm based on multiple hash tables have been proposed as a mode of indexing and then retrieval is performed based on Hough transform [5]. The authors have used minimal features like Short time Fourier transform and have tested for a total of 300 minute of audio input. The drawback of this approach is that it used a minimal feature set and hence did not a have a good precision and recall.

In another work by Shyamala Doraisamy et al, the authors have proposed a music information retrieval system based on N-gram technique which uses the presence of adjacent and concurrent patterns of the input segment [6]. The input signal which is a polyphonic music is used in its MIDI representation and using the concept of n-grams, overlap n-grams are constructed which are used for indexing and retrieval of the input music piece. Another N-gram based indexing technique was proposed for Indonesian Folk Songs by Aurora Marsye et al [7]. The authors have developed an algorithm to retrieve Indonesian folk songs which is based on pattern matching using the n-gram approach. The folksongs are indexed using the features like mfcc, and the indexing was done partially for some songs and fully for others. The fully indexed songs gave better results irrespective of the query length and the position of the query fragment.

Another work has been proposed by Yu-lung Lo et al for indexing and retrieval to address the problem of variable query length of the music segment during retrieval [8]. The authors have used existing indexing techniques in terms of suffix trees to match the query phrase into a tree representation and have addressed the varying query length of the segment.

In a work proposed by Yu-Lung Lo et al, the authors have used a concept called as Grid Suffix trees for the construct of indices which uses multi-feature for music indexing [9]. The grid suffix tree is conceptually visualized as a multi- dimensional suffix tree. The authors have claimed that this structure has a reduced space complexity when compared to other approaches.

In another grid-like structure proposed Chuan-Wang Chang et al, the representation is called dual-ternary indexing in which the authors have used a two-dimensional grid and three number notation to represent pitch content [2]. This work had a good space complexity but used a fixed segmentation algorithm and was not able to handle variable length query.

In general the various indexing techniques that are normally used for multimedia data are suffix tree, hash table, multi-dimensional suffix tree, N-gram model or some variations of a multi-dimensional representation. To that extent we have been motivated by a multi-dimensional representation and hash table concept. The concept of hash table motivated us because of the efficiency in retrieval. Therefore, the work done by Chuan-Wang Chang and Yu-lung Lo motivated us to go in for a multi-dimensional indexing structure to aid in a good time complexity even if compromises on space. In our work we have implemented the concept of multi-dimensionality using a modified Hash-table structure. The algorithm is based on the collision resolution of the hash table. This algorithm has been tested for Carnatic music songs and the performance compared with a modified dual-ternary indexing algorithm which is a variation to what is proposed by the authors indicated in [2] [10].

## 3. Characteristics of South Indian music

Indian Music is broadly divided into Hindustani music (North Indian music) and Carnatic music (South Indian counterpart). Both the systems are fundamentally similar and the differences arise in the style in which the notes are sung. Carnatic music can be thought of as a more systematized system as compared to Hindustani music [10]. The essential characteristics of Carnatic music are Ragam and Talam. Ragam is characterized by 'swaras'. There are essentially 7 swaras namely Sa, ri, ga, ma, pa, da, ni which is analogous to C, D, E, F, G, B, A in the keyboard. In an octave the 12 keys correspond to 12 frequencies. Normally, in a keyboard the C is played starting at 240 Hz. The next octave of C corresponds to a frequency of 480 Hz and therefore successive keys correspond to fixed frequencies, which follow a geometric progression. Carnatic music is based on a 22 interval per octave system as against Western music which is a 12 interval per octave system. Even if one chooses twelve keys to fill in an octave, there is no reason to tune them in a geometric progression. The seven swaras can be assigned frequencies between the frequency of the first chosen 'sa' and its harmonic frequency. The starting frequency of the 'Sa' is called the fundamental frequency and can vary between singers, between songs and a combination of singers and songs. Western music system does not have the concept of Raga as compared to Carnatic music. In addition, Western music is an even tempered system as against Carnatic music which is a just tempered system. Other characteristics of Carnatic music that are typically used for music information processing like Ragam, the variations of the swara, Talam, Gamakam are discussed by Rajeswari Sridhar et al [11]. Hence these specific characteristics of Carnatic music make it impossible to adapt the algorithms available for Western music directly to Carnatic music. Hence in our algorithm we have used the concept of converting the frequencies to swara representation which in turn is dependent on fundamental frequency of the input being considered and used this swara string as one of the feature for indexing. In the indexing algorithms proposed for Western music, note representation, fundamental frequency usage, multiple parameters and a combination of signal level features are not used for indexing. It was observed from literature that multiple features had given good result in some situations but those algorithms did not have a good indexing mechanism [4] [5]. Hence we have proposed an algorithm which uses this concept to generate a multi-feature based hashing algorithm which is discussed in the next section.

## 4. The Multi-Key Hashing Indexing Algorithm

This algorithm is based on HashTable. HashTable is a non-linear data structure which uses a key field among all fields from an element and this key is mapped on to an integer which is used as an index to store the element in the hash table. The key element is converted to an integer by means of a function called as hash function. This emphasizes the choice of a good hash function and a good key from the element for properly and uniquely mapping all the elements to a hash table. However, since the size of the hash table cannot be a large one and a unique hash function is impossible to achieve we end up in a situation called as collision. Collision is a concept in hashing in which more than one element's key map to the same index in a hash table. According to data structure literature the concept of collision is resolved using technique like chaining, re-hashing, etc. Chaining is the process in which, when multiple element's key map on to the same index we build a linked list of elements as a chain mapping to the same index value. But this could lead to a longer chain and hence a linear searching technique for one index while, the other index could be sparse. Re-hashing is another way of avoiding collision in which if collision occurs another hash function is used to compute the index and the element is stored in that index. In some cases, if the size of the hash table is small and the hash function is not sufficiently good enough then re-hashing would become endless. During the process of retrieval this hash function must be publicized to compute the index so that the data can be retrieved. In the case of chaining based collision resolution this would lead to a linear search on a particular index value. In the case of re-hashing all of the available hash functions are to be publicized and the sequence of usage of hash function must also be specified. Hence, we have proposed an algorithm that uses a combination of chaining and a variation of re-hashing and we call it as multi-key hashing. The already proposed algorithms for western music had given better results when more signal level features were used. In addition, we would like to use the multi-dimensional grid structure for efficient storage which motivated us to propose this algorithm.

Multi-key hashing is similar to hashing in which we use a key to compute the index using a hash function and store this element in the hash table. In the event of collision we first chain the element in the index and we use the next key of the element and use the same hash function to compute the index. The element is stored in the next index if there is no collision. If collision occurs then we use the next key to compute the index. The order in which the key should be chosen from the list of available keys is determined based on their uniqueness and arranged in decreasing order of robustness. Thus the key which is most robust is ranked first followed by the next robust key and so on. During retrieval the same process is carried out and this justifies the use of only one hash function and at the same time duplication is allowed by the process of chaining. This concept is represented in Figure 1.

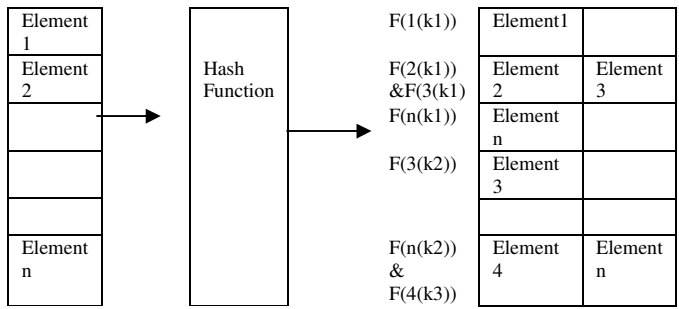

**Figure 1: Multi-key Hashing**

In the above representation, Element1, Element 2, Element 3 corresponds to the available 'n' elements that need to be indexed. Hash function is unique function that uses the keys to compute the index. Let us represent 1(k1) as the $1^{st}$ key of the $1^{st}$ element, 1(k2) corresponds to the $2^{nd}$ key of the $1^{st}$ element and so on. Using the key 1 the index is computed. If there is no collision the element is added to the hash table. In the event of collision we chain the element and then use the same hash function to compute the index using another key. This is illustrated in Figure 1. F(1(k1)) indicates that the hash function F( ) is computed using key 1 of the $1^{st}$ element which is actually mapped onto an index value.

This algorithm is justified for multi-media indexing and retrieval since a multi-media piece is characterized by more than one media and each of the media can be one key. In our work we have considered this algorithm for indexing songs. A music song is characterized by various parameters of Rhythm, pitch, melody, tempo, etc. Each of these parameters is represented using the signal characteristics of Frequency, Spectral Flux, Spectral Centroid, Mel-frequency cepstral coefficients, Zero-crossings and other features. For our work we have used the spectral parameters, in which Frequency converted to a melody swara representation, Spectral flux, Spectral Centroid and Mel-frequency cepstral coefficients (MFCC). Spectral centroid refers to the median of the spectrum. It is computed by estimating the power spectrum of the signal [12]. Spectral flux refers to the change in the spectral energy between successive samples [12]. MFCC is computed as a spectrum of spectrum. These coefficients which were developed for speech processing are being used as one of the key feature for signal processing applications which is defined as the log magnitude of the power of the signal and this is the next parameter which we have used for indexing. The other feature is the melody string comprising of S, R, G, M, P, D, N swara from the input song. To derive the swaras from the input song, the fundamental frequency of the input signal is required to determine the swara pattern. This fundamental frequency is computed using the YIN algorithm by Alain de Chevigne, et ak [13]. Using the fundamental frequency and the extracted frequency components from each segment of the input sequence a ratio is computed and this ratio is mapped onto a swara sequence of S, R, G, M, P, D, and N [10]. These four features namely the swara sequence constituting the melody string, spectral flux, spectral centroid and mfcc are used for indexing the music piece. Later these features in the same order are used to compute the index of the hash table and retrieval is performed.

The algorithm is summarized in Figure 2.

*Multi-keyHashing (Input elements n)*
{
For elements = 1 to n
Extract-feature(element[i])
Features like melody string, flux, centroid and mfcc
For elements = 1 to n

```
{
        Index = Hashfunction(Melody seq)
        If(!Collision)
                Store element in the Index
                Go to next element
        Else
                Chain(element)
                Index = Hashfunction(flux)
                If(! Collision)
                        Store element in the Index
                        Go to next element
                Else
                   Chain(element)
                   Index = Hashfunction(centroid)
                   If(!Collision)
                        Store element in the Index
                        Go to next element
                  Else
                      Chain(element)
                     Index = Hashfunction(MFCC)
                     If(!Collision)
                        Store element in the Index
                        Go to next element
}
```

Figure 2: Algorithm Multi-key indexing

## 5. Modification to the existing western music algorithm

In the dual ternary indexing approach proposed by Chuan-Wang Chang et al [2] , the songs stored in the database are segmented into meaningful phrases and indexed. This approach uses a two-dimensional array for indexing and the two indices that are used are called as $I_{PRE}$ and $I_{SUF}$. $I_{PRE}$ and $I_{SUF}$ are calculated from prefix and suffix part of the phrase contour as follows.

$$I_{PRE} = f(P_{PRE}) = \sum_{i=0}^{i=L_{I_{PRE}}-1} PC_{L_P-1-i} \times 3^{L_{I_{PRE}}-1-i}$$

$$I_{SUF} = f(P_{SUF}) = \sum_{i=0}^{i=L_{I_{SUF}}-1} PC_i \times 3^{L_{I_{SUF}}-1-i}$$

The total length of the index is fixed much prior to the implementation. When collision occurs, it is resolved by open chaining. The algorithm computes a phrase contours which hash to same indices are stored in the same slot of the indexing structure. Figure.3 shows the dual ternary indexing structure. The phrase contour consists of the values '0', '1' and '2'. A '0' is assigned to a segment if the frequency of the current segment is the same as the previous one, a '1' is assigned if the frequency of the current segment is less than the previous segment and a '2' is assigned if the frequency of the current segment is greater than the previous segment. Hence the possible values for a phrase contour consist of 0, 1, and 2. We have modified the algorithm of dual ternary indexing in the way this contour is determined. We initially determine the fundamental frequency of the signal using the algorithm specified in [13]. Then we determine the dominant frequency of every small segment by segmenting it and using the fundamental frequency and the frequency of every segment we determine the swara sequence S, R, G, M, P, D, N. After determining the swara sequence we take the phrase and convert the change in swara using the phrase representation of '0', '1' or '2' by identifying the change in swara between successive notes. In the figure 3, phrase1 and phrase2 may have same contour but may have a different melody string conveying different raga or belong to different music file.

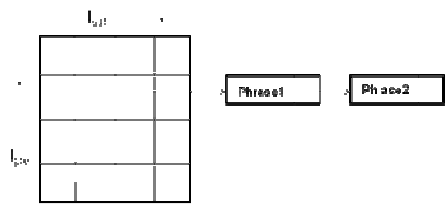

Figure.3. Dual Ternary Indexing Structure

In addition to the modification done during indexing we have modified the algorithm during query processing. During query processing, the aim of exact matching was to retrieve all phrases that have both the same prefix and suffix as those of the query. In this case, the music piece can be easily retrieved irrespective of the index length by calculating $I_{PRE}$ and $I_{SUF}$. The aim of similarity matching was to retrieve all phrases that had either the same prefix or suffix as that of the query but differs in length. If the query is the prefix part of the phrase, then suffix query expansion is done to generate all possible queries that have the same prefix as the input query. Similarly, if the query is the suffix part of the phrase, then prefix query expansion is done. Thus entries in the same row or column are retrieved for similarity matching. In the dual ternary index construction module, songs stored in the database are segmented into meaningful phrases. The algorithm is based on the assumption that users tend to submit a perceptually meaningful query phrase. It uses representative music fragment extraction technique [14] to extract representative melody fragments. Hence this algorithm will not be able to handle all segmentation techniques since other segmentation techniques will not be able to segment in such a manner to have the segment having either prefix or suffix of a melody contour stored in the indexing table. For example, if stored melody contour S = 2 2 0 0 1 2 and Query Q = 2 0 0, then S will not be retrieved though it contains Q. In order to avoid this we modify this existing algorithm to use a sliding technique and construct the index structure. The concept of using sliding window has been adopted by us from C.W. Chang et al, [15]. Hence this process of sliding each phrase during construction of the indexing structure and storing requires large memory. So we slide the query during query processing by using a fixed window to allow the query to be any part of the melody contour. If N is the length of melody contour and W is the window size, then number of sub-segments is N-W+1. For example, if N = 8 and W = 4, then Number of sub-segments = 8-4+1 = 5. Since we have considered this algorithm for Carnatic music we have using the sliding window size of 7 to handle variable query length. Using this sliding window technique we index the songs and later retrieval is performed in a similar manner as done by the original algorithm of dual ternary indexing.

## 6. Experimental Set up and Evaluation

In this work we take input music songs from Tamil film music and Carnatic music sung by singers like Nithyasree Mahadevan, M.S. Subbulakshmi, Ilayaraja, Balamuralikrishna. The input songs are polyphonic and hence consist of voice and accompanying instruments. Since we are using the fundamental frequency algorithm to estimate the melody string the input signal is made to run through a signal separation algorithm [11] to separate the voice and music instruments. The voice signal alone is taken for processing. This voice signal is sampled at 44.1 KHz and then we segment it using the Talam characteristics of the input signal [11]. Then the segmented signal is used to determine the frequency components of the signal from which we estimate the frequencies present in the signal. The fundamental frequency of the signal is also estimated using the YIN algorithm and using the estimated frequency value of every segment and the fundamental frequency the swara pattern is estimated for the input signal [11]. This swara pattern forms the melody string feature to be used for indexing. In addition to the above we also estimate the frequency based phrase contour by comparing the frequency segments of successive frames to be used by the Dual ternary indexing algorithm. In addition, other features extracted are MFCC, Spectral flux and Spectral Centroid. These features along with the melody string are used for indexing using the multi key hashing algorithm.

To implement the modified dual ternary indexing algorithm we used the segmentation as proposed for Carnatic music and also used the sliding window technique. We used as sliding window size of 7 which is the size of the length of the swara in a Parent raga [10]. The representation of the melody sequence is similar to the original algorithm as '0', '1', '2' indicating change or no change between successive segments

## Table 1. Indexing and Retrieval Observations

| S.No | Indexing algorithm used | Average Time Efficiency for similarity matching (seconds) | Average Time efficiency for Exact matching (seconds) | Space Effic-iency (KB) |
|------|------------------------|-----------------------------------------------------------|------------------------------------------------------|------------------------|
| 1 | Dual – ternary with modification | 4.568700 | 3.718300 | 333 (5_5) |
| 2 | Multi-key Hashing | 3.857275 | 0.049131 | 1280 KB |

A sample of 100 songs belonging to all the singers indicated is taken for training and these songs are indexed in two different databases one using modified dual ternary indexing and other using multi-key hashing separately in each database. The algorithm is implemented as earlier after extracting features, determination of hash index value, computing phrase contour adopting a sliding window technique and the database in populated with the indices in both the databases. Then we tested with another sample set of 100 songs and also used the trained 100 songs for testing. The observations are tabulated in Table 1.The observations are tabulated based upon choosing songs belonging to four singers in 5 different ragas and pertaining to varying genre. The characteristics of raga, genre or singer are not identified but the features conveying them are being used for indexing. From Table 1 it is clear that multi-key hashing has a faster time efficiency for retrieval than modified dual-ternary indexing when an empirical analysis is done on a data set consisting of 200 songs. However the space efficiency of the dual-ternary indexing algorithm is better compared to multi-key hashing algorithm. For modified dual ternary indexing algorithm we implemented a 2-dimensional table of 5 * 5 and we observed that on an average the space complexity was 333 KB. On the other hand, since multi-key hashing saves multiple copies of the same element in different positions had a higher space complexity. However the time-complexity of multi-key hashing is better because in modified dual-ternary indexing the concept of computing the phrase contour is present which consumes the additional time. Considering the availability of hardware we can compromise space over time. Hence we suggest that our algorithm is better than the modified dual-ternary indexing algorithm. We cannot adopt dual ternary indexing algorithm without modification since it requires a pre-defined segmentation algorithm and did not handle any query length.

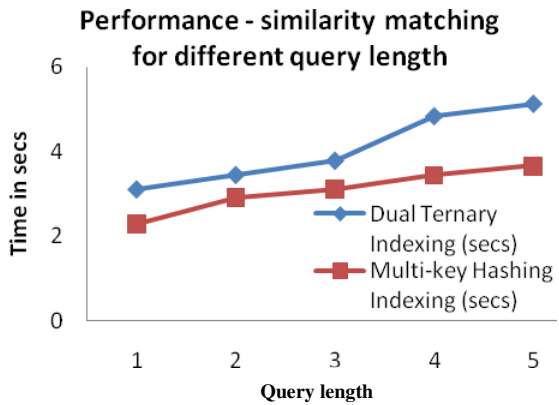

**Figure 4: Performance for different query length for similar matching**

We also analyzed the performance of this algorithm for different query length and the following observations were made in terms of average time elapsed which is given in Figure 4

The same performance analysis on varying query lengths are performed for exact matching and the observations are discussed in Figure 5

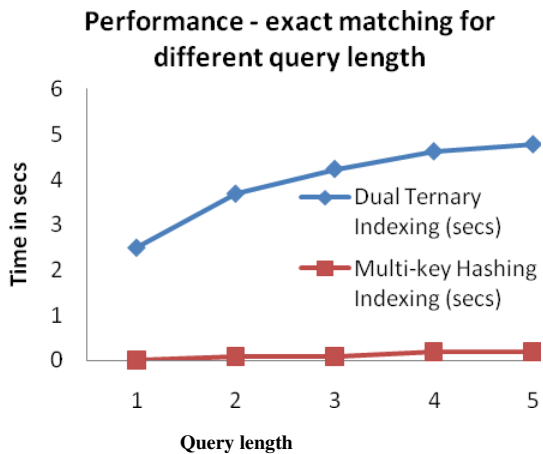

**Figure 5: Performance for different query length for exact matching**

It can be observed that for similarity time matching the time taken for retrieval is almost same for dual-ternary indexing and multi-key hashing whereas for exact matching the time taken by dual ternary indexing is much higher than that for multi-key indexing.

We then estimated Precision and recall measure to test the performance of the algorithm. The results are tabulated in Table 2.

**Table2: Precision and Recall**

| Parameter | Modified Dual Ternary Indexing | Multi-key Hashing Indexing |
|---|---|---|
| Precision | 70% | 65% |
| Recall | 68% | 70% |

The precision rate of multi-key hashing is lower than that of modified dual ternary indexing. This difference is because of the availability of one element in multiple positions and hence the number of songs retrieved is more than what is actually should be retrieved. On the other hand the recall of multi-key indexing is high again because of the presence of element at multiple locations and hence whatever should be retrieved was retrieved.

The comparison of the two algorithms is given in Table 3

**Table3: Overall Comparison of the algorithms**

| Modified Dual-Ternary Indexing | Multi-Key Hashing |
|---|---|
| Uses contour representation of the melody string as the key for indexing | Uses melody string, MFCC, Flux and Centroid musical features as keys for indexing |
| Resolves collision by chaining | Resolves collision by chaining and duplication |
| Compromises time over space complexity | Compromises space over time complexity |
| Suitable for Western music | Suitable for Western music and Carnatic music |
| Works only when song is segmented into meaningful phrases and people give meaningful query phrase | Works for any segmentation technique used |

## 7. Conclusion

In this work we have compared the performances of two indexing algorithms, the one that we have proposed and the one that is already available for indexing songs and have performed retrieval of the same using the same indexing algorithms. We analysed the algorithm based on their time and space complexity using empirical analysis. We also analysed the algorithm for their precision and recall measure as a method to estimate performance. This algorithm could be improved by using high level models for indexing and determining a known pattern sequence. Comparisons could also be made with other known indexing algorithms for their time and space efficiency to conclude on the appropriate algorithm for music information retrieval irrespective of the type of music